\documentclass[amsmath,reprint]{revtex4-1}

\usepackage{graphicx}
\begin{document}

\title
{Gate-controlled generation of optical pulse trains using individual carbon nanotubes}
\author{M.~Jiang}
\author{Y.~Kumamoto}
\author{A.~Ishii}
\author{M.~Yoshida}
\author{T.~Shimada}
\author{Y.~K.~Kato}
\email[Corresponding author: ]{ykato@sogo.t.u-tokyo.ac.jp}
\affiliation{Institute of Engineering Innovation, 
The University of Tokyo, Tokyo 113-8656, Japan}

\begin{abstract}
We report on optical pulse-train generation from individual air-suspended carbon nanotubes under an application of square-wave gate voltages. Electrostatically-induced carrier accummulation quenches photoluminescence, while a voltage sign reversal purges those carriers, resetting the nanotubes to become luminescent temporarily.
Frequency domain measurements reveal photoluminescence recovery with characteristic frequencies that increase with excitation laser power, showing that photoexcited carriers quench the emission in a self-limiting manner. Time-resolved measurements directly confirm the presence of an optical pulse train sychronized to the gate voltage signal, and flexible control over pulse timing and duration is demonstrated.
\end{abstract}

\maketitle

Unique excitonic processes in single-walled carbon nanotubes result in complex photocurrent and electroluminescence mechanisms \cite{Gabor:2009, Barkelid:2014, Mann:2007, Mueller:2010}, offering diverse opportunities in nanoscale optoelectronics. In particular, interplay between free carriers and excitons  \cite{Perebeinos:2008, Kinder:2008, Steiner:2009, Matsuda:2010, Yasukochi:2011} is known to play an important role in determining the efficiencies in such devices, but its dynamical response has remained elusive. 

Here we investigate the luminescence response of individual carbon nanotubes subjected to square-wave gate voltages, and unexpectedly find that such exciton-carrier interactions result in optical pulse-train generation. By performing experiments in both frequency and time domains, we show that the voltage transitions can temporarily purge photocarriers to generate self-limiting optical pulses. Our results demonstrate flexible control over pulse timings and widths, expanding the possibilities for optoelectronic circuits using carbon nanotubes \cite{Kim:2014}.

\begin{center}
\begin{figure}[h]
\includegraphics{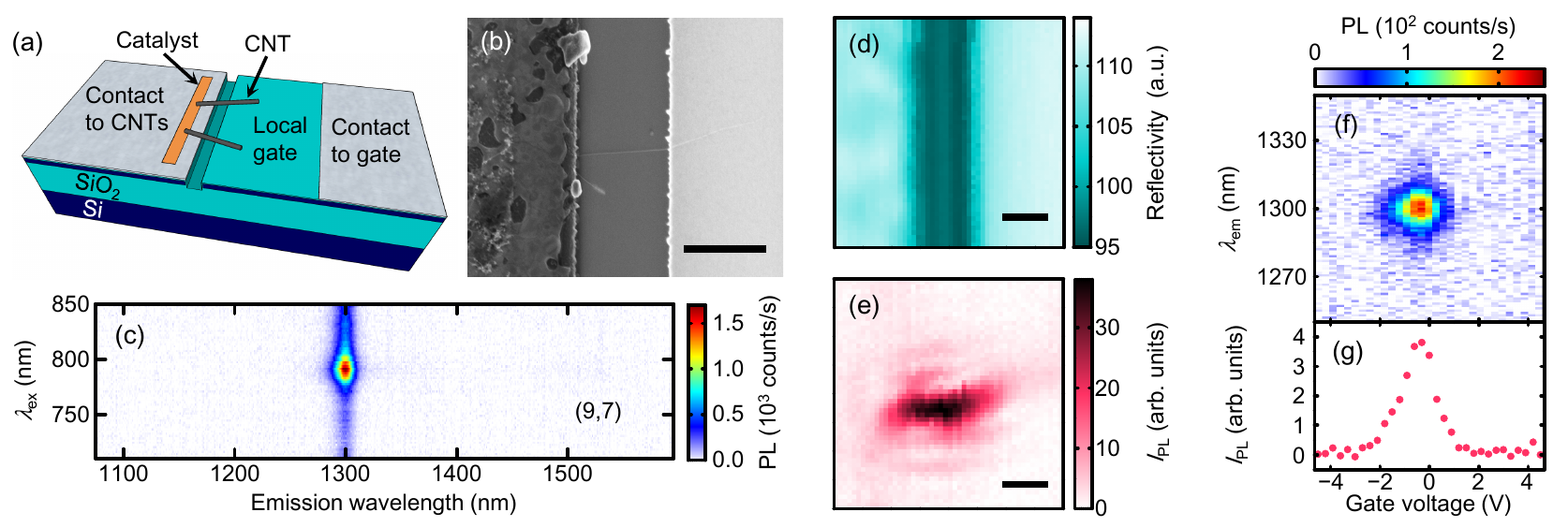}
\onecolumngrid
\caption{\label{fig1}
Device characterization.
(a) Schematic of a device.
(b) Electron micrograph of a device. 
(c) PL excitation map of a nanotube in a device measured in (d-g).
(d) Reflectivity image taken with $\lambda_\text{ex}=761$~nm. 
(e) PL image of the same area as (d). 
(f) PL spectra as a function of gate voltage, taken at $P=1$~$\mu$W.
(g) Gate voltage dependence of $I_\text{PL}$. 
In (b, d, e), scale bars are  1~$\mu$m. (c-e) are taken with $P=20$~$\mu$W and (e-g) are taken at $\lambda_\text{ex}=790$~nm. The spectral integration window for $I_\text{PL}$ is from $\lambda_\text{em}=1275$~nm to 1325~nm. 
}\end{figure}
\end{center}

A schematic of our device is shown in Fig.~\ref{fig1}(a). Air-suspended carbon nanotubes (CNTs) are contacted on one side of a trench, and a local gate on the opposite side is used for applying electric fields onto the CNTs. By performing nanotube growth at the last step, we are able to take advantage of the superior optical properties of as-grown suspended nanotubes  \cite{Lefebvre:2006, Hofmann:2013, Sarpkaya:2013}, while the use of the top silicon layer of a silicon-on-insulator substrate allows for efficient and fast gating.  

The devices are fabricated in a manner similar to suspended-CNT field-effect transistors \cite{Yasukochi:2011, Kumamoto:2014}, but with silicon-on-insulator substrates with 260~nm of top silicon layer and 1~$\mu$m of buried oxide. We start by dry etching trenches through the top silicon layer, followed by a wet etch to further remove $\sim 200$~nm of buried oxide. The top silicon layer is thermally oxidized at $900^\circ$C for one hour to form a SiO$_2$ layer with a nominal thickness of 20~nm. The nanotube contact is placed right next to the trench, while the contact for the gate is located $5$ or 10~$\mu$m away from the trench. For the electrodes, we evaporate 2~nm of Ti followed by 30~nm of Pt. From catalyst particles placed on the electrodes, nanotubes are grown over the trench onto the gate by chemical vapor deposition \cite{Maruyama:2002, Imamura:2013}. An electron micrograph of a nanotube in a device is shown in Fig.~\ref{fig1}(b).

\onecolumngrid
\twocolumngrid

We characterize the emission properties of devices using a confocal microspectroscopy system similar to that described in previous work \cite{Moritsubo:2010, Watahiki:2012}, but with an automated three-dimensional stage for scanning the sample instead of a laser scanning mirror. The samples are excited with a wavelength-tunable continuous-wave Ti:sapphire laser, and PL is detected by an InGaAs photodiode array attached to a spectrometer. The laser polarization angle is adjusted to maximize the PL signal, and excitation wavelength is tuned to the $E_{22}$ resonance unless otherwise noted. All measurements are done in air at room temperature. 

Figure~\ref{fig1}(c) shows PL as a function of excitation wavelength $\lambda_\text{ex}$ and emission wavelength $\lambda_\text{em}$ taken with a laser power $P=20$~$\mu$W, from which the chirality (9,7) is assigned \cite{Bachilo:2002, Ohno:2006prb}. By comparing the reflectivity image [Fig.~\ref{fig1}(d)] to the PL image obtained by mapping out the integrated PL intensity $I_\text{PL}$ [Fig.~\ref{fig1}(e)], we confirm that the tube is fully suspended. The dc gate-voltage characteristics of the device is shown in Figs.~\ref{fig1}(f) and \ref{fig1}(g). Upon application of a gate voltage $V_\text{g}$, PL quenching occurs as a result of phase-space filling and doping-induced exciton relaxation \cite{Perebeinos:2008, Kinder:2008, Steiner:2009, Matsuda:2010, Yasukochi:2011}.

\begin{figure}
\includegraphics{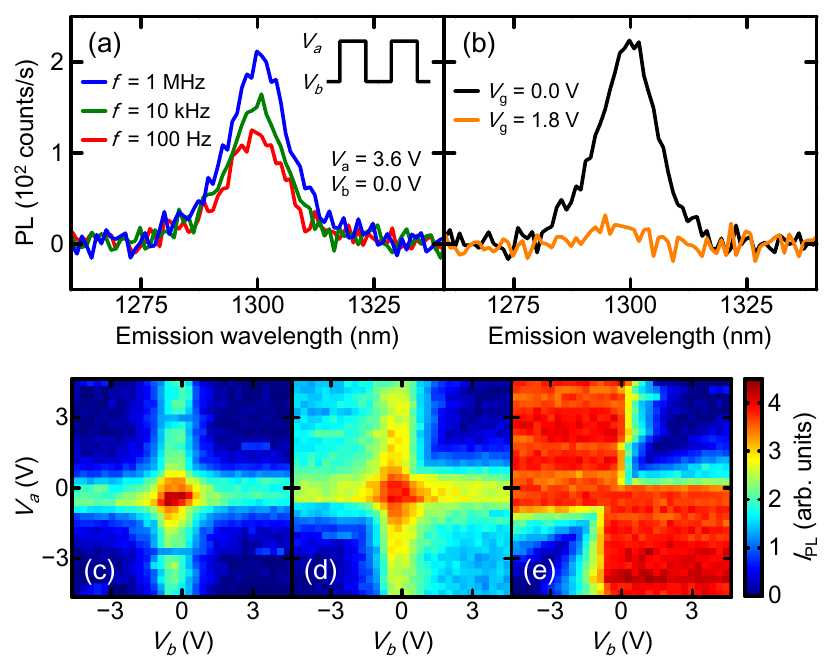}
\caption{\label{fig2}
Square-wave gate-voltage induced PL recovery.  
The device characterized in Fig.~\ref{fig1}(c-g) is measured with $P=1$~$\mu$W.
(a) PL spectra taken under square-wave voltage with $V_a=3.6$~V and $V_b=0.0$~V at $f=100$~Hz (red), 10~kHz (green), and 1~MHz (blue). Inset is a schematic showing the definitions of $V_a$ and $V_b$.
(b) PL spectra taken with dc voltages of  $V_\text{g}=0.0$~V (black) and  $V_\text{g}=1.8$~V (orange).
(c-e) Integrated PL as a function of $V_a$ and $V_b$ at (c) $f=100$~Hz, (d) 10~kHz, and (e) 1~MHz.   The PL integration window is from $\lambda_\text{em}=1275$~nm to 1325~nm. 
}\end{figure}

Interestingly, we find that such quenching can be eliminated at high frequencies when square-wave gate voltages are applied. As shown in the inset of Fig.~\ref{fig2}(a), we use a waveform that alternates between $V_a$ and $V_b$ at a frequency $f$, with a rise/fall time of 20~ns. The PL spectra taken with $V_a=3.6$~V and $V_b=0.0$~V for various $f$ are shown in  Fig.~\ref{fig2}(a). At $f=100$~Hz, the PL intensity is about half the intensity of that taken at $V_\text{g}=0.0$~V [Fig.~\ref{fig2}(b), black curve]. This is expected as PL is quenched for half of the time. As the square-wave frequency becomes higher, however, the emission intensity increases, and it recovers to the zero-voltage level at $f=1$~MHz [Fig.~\ref{fig2}(a), blue curve]. 

The observed PL recovery cannot be due to time-averaging of the gate voltage at high frequencies, as the capacitive cut-off frequency of the device is estimated to be larger than $50$~MHz. In fact, we observe significant quenching for a static field corresponding to the time-averaged voltage [Fig.~\ref{fig2}(b), orange curve].

In order to investigate the mechanism underlying the PL recovery, we have measured the integrated PL intensity as a function of  $V_a$ and $V_b$ for three different frequencies. At $f=100$~Hz [Fig.~\ref{fig2}(c)], the PL intensity behaves as expected from the dc characteristics. PL is brightest when  $V_a =V_b =0$~V, and quenching is observed when both $V_a$ and $V_b$ are non-zero. Along the lines that correspond to $V_a=0$~V and $V_b=0$~V, the PL intensity is approximately half of that for $V_a =V_b =0$~V, resulting in the cross shape in the  $V_a$-$V_b$ map.

As the frequency is increased to $f=10$~kHz, we find that the PL starts to recover in the top-left and bottom-right areas in the map [Fig.~\ref{fig2}(d)]. These areas correspond to the cases where the signs of $V_a$ and $V_b$ are opposite, suggesting that PL recovery takes place through a process occurring at a voltage transition passing through 0~V. At $f=1$~MHz, the PL intensity in those two areas in the $V_a$-$V_b$ map becomes uniform, indicating that the recovery is complete [Fig.~\ref{fig2}(e)].

\begin{figure}
\includegraphics{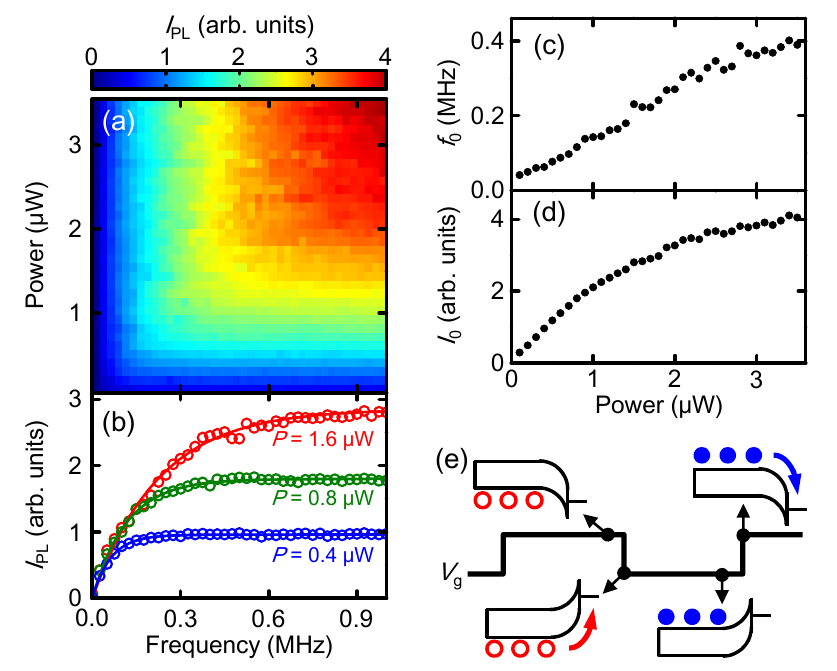}
\caption{\label{fig3}
Frequency domain measurements.
(a) Integrated PL as a function of $f$ and $P$ for a device with a (10,8) nanotube. $V_a=-V_b=3.0$~V, and $I_\text{PL}$ is obtained by integrating the PL spectra from $\lambda_\text{em}=1400$~nm to 1450~nm. 
(b) Frequency dependence of integrated PL for $P=0.4$~$\mu$W (blue), $P=0.8$~$\mu$W (red), and $P=1.6$~$\mu$W (blue).
(c) and (d) Power dependence of $f_0$ and $I_0$, respectively.
(e) Schematic showing a microscopic physical mechanism. Red open circles and blue filled circles represent holes and electrons, respectively.
}\end{figure}

We further examine the PL recovery process in the frequency domain. In Fig.~\ref{fig3}(a), PL intensity is plotted as a function of the square-wave frequency and the excitation laser power for another device. Typical frequency dependence data are shown in Fig.~\ref{fig3}(b) for three different excitation powers. $I_\text{PL}$ recovers linearly at low frequencies and shows saturation at high frequencies, consistent with the observation in Fig.~\ref{fig2}. It is noteworthy that $I_\text{PL}$ does not depend on the laser excitation power in the low frequency regime, which implies that the emission intensity is limited by the number of voltage transitions. In addition, it can be seen that the saturation occurs at a higher frequency when the excitation power is increased.

In order to extract the characteristic saturation frequency $f_0$, we fit the data to $I_\text{PL}=I_0 [1-\exp(-f/f_0)]$, where $I_0$ is the saturation intensity. As shown in Fig.~\ref{fig3}(b), we obtain good fits, and the results are summarized in Figs.~\ref{fig3}(c) and \ref{fig3}(d). The saturation frequency increases linearly with laser power, indicating that the photoexcited carriers are playing a role in the PL recovery process. The saturation intensity, which would be equivalent to the zero-voltage intensity, shows a sublinear behavior known to be caused by efficient exciton-exciton annihilation in CNTs \cite{Wang:2004, Matsuda:2008, Murakami:2009prl, Xiao:2010, Moritsubo:2010}.

In Fig.~\ref{fig3}(e), we propose a model that can explain the experimental observations. Just before a transition from positive to negative gate voltage, holes have accumulated in the nanotube. In such a state, PL is efficiently quenched by the Auger process involving the carriers \cite{Perebeinos:2008, Kinder:2008, Steiner:2009, Matsuda:2010, Yasukochi:2011}. When the voltage is reversed, the holes are easily swept into the contact as there exists no potential barriers. Since the tube is now free of carriers, it becomes luminescent until the photoexcited electrons have accumulated enough to quench the PL in a self-limiting manner. Essentially, the voltage steps reset the nanotubes to become bright again. For the opposite polarity voltage transitions, the nanotube should also become temporarily luminescent because of the conduction and valence band symmetry \cite{Jarillo-Herrero:2004}.

This model explains all of the key experimental features observed. PL recovery only occurs when the voltage transition passes through 0~V, as such a voltage sign reversal is required to sweep the carriers into the contact. Because nanotubes become luminescent for every voltage step, PL intensity is proportional to $f$ in the low frequency regime. At high-frequencies, the reset happens before the carriers are accumulated, suppressing the quenching and recovering the PL emission to the zero-voltage level. Since carrier accumulation happens faster for higher powers, the linear power dependence of the saturation frequency can also be explained.

\begin{figure}
\includegraphics{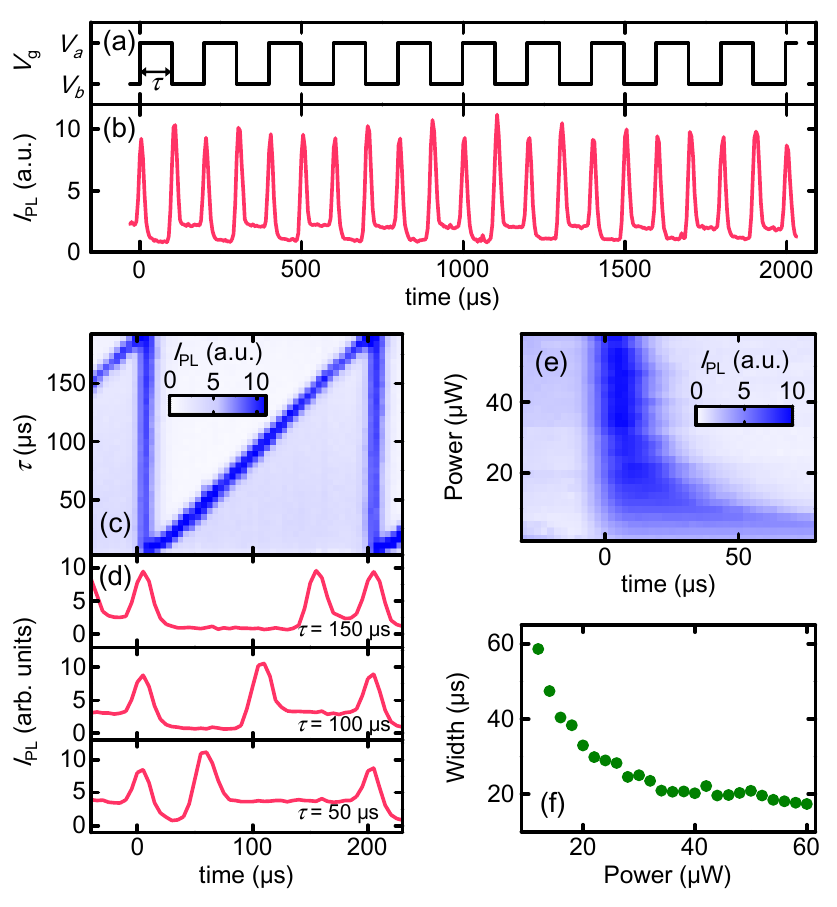}
\caption{\label{fig4}
Time-domain measurements.
A (10,8) nanotube is measured with $V_a=+1.2$~V, $V_b=-1.8$~V, and $f=5011$~Hz. The spectral integration window for $I_\text{PL}$ is from $\lambda_\text{em}=1440$~nm to 1490~nm.  
(a) and (b) Time dependence of gate voltage and integrated PL, respectively.  
(c) Integrated PL as a function of time and $\tau$.
(d) Temporal evolution of integrated PL for $\tau=50$, 100, and 150~$\mu$s.
For (b-d), $P=50$~$\mu$W is used.
(e) Integrated PL as a function of time and $P$.
(f) Laser power dependence of pulse width.
}\end{figure}

If this model were correct, we expect pulsed light emission to occur just after the voltage transitions. We have performed time-domain measurements of the spectrally integrated PL on another device to directly observe such optical pulses. An optical chopper phase-locked to the function generator is placed at the entrance of the spectrometer. The chopper has a duty cycle of approximately 6\% giving a temporal resolution of $\sim13$~$\mu$s at $f=5011$~Hz, and the phase relative to the square-wave is scanned to obtain the time dependence. The results are shown in Figs.~\ref{fig4}(a) and \ref{fig4}(b), unambiguously confirming the existence of a series of pulses that are synchronized with the gate voltage transitions. The device effectively converts an electrical clock into an optical pulse train with twice the frequency, consistent with the microscopic physical mechanism proposed above. 

As the optical pulses are generated by the voltage transitions, pulse timings can easily be controlled through the gate voltage waveform. In Figs.~\ref{fig4}(c) and \ref{fig4}(d), we present measurements performed for various time delay $\tau$ between the upward and the downward voltage steps. The pulse timing can be tuned continuously throughout the repetition period, suggesting that more complex sequences would be possible.

In addition to the timing, it is also possible to control the temporal width of the pulses through the excitation laser power. Because the pulse duration is determined by the carrier accumulation time, we expect pulse width to become narrower at higher powers. Such a control is shown in Fig.~\ref{fig4}(e), in which the power dependence of the PL intensity near the voltage step is plotted. At low powers, the pulse width is broad and the PL intensity is low. As the power is increased, the PL intensity becomes stronger, and at the same time, the pulse width shortens significantly. We plot the full-width at half-maximum of the PL pulses in Fig.~\ref{fig4}(f), showing pulse narrowing to $\sim17$~$\mu$s which is limited by the time resolution of our measurement setup.

Such a change in the pulse width explains why the PL intensity does not depend on excitation power at low frequencies [Figs.~\ref{fig3}(a) and \ref{fig3}(b)]. If we assume that the emission rate is proportional to the excitation power, the time-integrated emission intensity per pulse should be independent of the laser power because the accumulation time is inversely proportional to the laser power, cancelling out the power dependence.

Our results demonstrate the potential for performing electrical-to-optical signal conversions at the nanoscale using individual nanotubes. A unique feature of the pulse train generation presented here is that only a voltage step, rather than a voltage pulse, is needed to generate an optical pulse, implying that the optical pulse train can have a higher bandwidth than the electrical signal. In principle it should be possible to generate much shorter pulses if we place the nanotubes in vacuum, as it allows the excitation power to be increased by a few orders of magnitude. 

\begin{acknowledgments}
Work supported by KAKENHI (24340066, 24654084, 26610080), SCOPE, Canon Foundation, Asahi Glass Foundation, and KDDI Foundation, as well as the Nanotechnology Platform and Photon Frontier Network Program of MEXT, Japan. 
\end{acknowledgments}

\bibliography{Pulsetrain}

\end{document}